# LensMEM: A Gravitational Lens Inversion Algorithm Using the Maximum Entropy Method

Sylvanie Wallington, Christopher S. Kochanek, & Ramesh Narayan

Harvard-Smithsonian Center for Astrophysics 60 Garden Street Cambridge, MA 02138

## ABSTRACT

We present a new algorithm for inverting poorly resolved gravitational lens systems using the maximum entropy method (MEM). We test the method with simulations and then apply it to an 8 GHz VLA map of the radio ring lens MG1654+134. We model the lens as a singular isothermal sphere embedded in an external shear field and find the critical radius of the lens is $b = 0\rlap{.}''9820$, the dimensionless shear is $\gamma = 0.0771$, and the position angle of the shear is $\theta = 100\rlap{.}°8$. These results are consistent with the results obtained by Kochanek (1995) using a complementary inversion algorithm based on Clean.

## 1. Introduction

Since the discovery of the first gravitational lens (Walsh et al. 1979), the number of observed lens candidates has grown enormously (see Keeton & Kochanek 1995 for a review of the data). Today there are dozens of well-established lens candidates observed at various wavelengths and having a range of morphologies. These include multiply-imaged quasars, radio rings, giant luminous arcs, and arclets. These huge observational advances drive theoretical efforts to exploit lensing as an astrophysical tool. Lensing has many applications: it provides an independent method for measuring the Hubble constant (Refsdal 1966a,b, Falco, Gorenstein & Shapiro 1991, Narayan 1991), it allows a direct probe of the gravitational effects of dark matter (Paczyński 1986, Narayan, Blandford & Nityananda 1984, Kaiser & Squires 1993), it provides us with a highly magnified view of some of the most distant objects in the universe (Kochanek et al. 1989), and it probes the cosmological model (Turner 1990, Kochanek 1995). Some applications involve the statistical analysis of the entire lens sample, but many applications require a detailed understanding of an individual lens system. This has made lens modeling an increasingly important field.

Ring and arc lenses are particularly well suited to detailed modeling. Well-resolved images of these objects provide many more constraints for the model source and lens potential than point lenses such as multiply-imaged quasars. The constraints created by multiple imaging allow the unknown source and lens to be solved for simultaneously. Sophisticated lens modeling algorithms have been developed in recent years as the number of observed lenses has grown. In a rigorous modeling scheme, no *a priori* source structure is assumed. Generally, a model is constructed for the lensing mass or potential and the image is "inverted" by mapping through the lens back to the source plane. The lens parameters are modified until they produce the best fit of the inversion to the data. In the most sophisticated inversion algorithms, such as LensClean (Kochanek & Narayan 1992), the standard inversion procedure is combined with an algorithm that compensates for the finite resolution of the observations during the reconstruction.

We have developed a new algorithm, LensMEM, for inversion of extended gravitational lens images. LensMEM is similar to LensClean, but uses the maximum entropy method, MEM, rather than the Clean algorithm to correct for the smearing of the data by finite resolution. In regular astrophysical applications the two methods are complementary: Clean is better suited for concentrated, point-like sources, while MEM is better at deconvolving extended sources with low level structure. In lensing applications, both types of deconvolution algorithms are valuable.

In an early paper (Wallington, Narayan & Kochanek 1994, hereafter WNK), we developed the LensMEM method in one dimension in order to work out the details of the algorithm on a manageably sized problem. In a separate paper (Wallington, Kochanek & Koo 1995, hereafter WKK), we developed a two-dimensional mapping algorithm which lacked any correction for finite resolution, but allowed us to produce an accurate inversion in the absence of significant atmospheric distortion. The inversion scheme using this mapping method was applied to the giant arc in Cl0024+1654, for which relatively high resolution images made beam-deconvolution less necessary. In this paper we expand the MEM equations to two dimensions and combine them with the mapping method from WKK to produce a full-fledged, two-dimensional MEM lens reconstruction algorithm.

In §2 we describe the details of the pixel mapping, MEM equations, and solution method. We demonstrate the scope and accuracy of the algorithm in §3 by testing it on artificial data for which we know the true lens and source. Finally, in §4 we apply LensMEM to real data from the radio ring MG1654+134. In §5 we provide a summary.

## 2. Method

The algorithm works in two cycles. The inner cycle finds the best-fit source model for the current, fixed, lens model, always starting from the same initial conditions. The residuals from this fit are an estimate of how well the lens model fits the data. In the outer cycle the parameters of the lens model are adjusted to minimize the residuals.

This general procedure was originally developed by Kochanek et al. (1989) with a crude inner cycle that took no account of the finite resolution of the observations. Here we combine the more advanced mapping algorithm of WKK with the MEM methods developed in WNK to build an MEM-based inner cycle. The alternate inner cycle is the LensClean method developed by Kochanek & Narayan (1992) and Ellithorpe, Kochanek & Hewitt (1995). The general form of these algorithms differs only in the image reconstruction method used for the inner cycle.

## 2.1. Pixel Mapping Technique

The source position, $\mathbf{u}$, corresponding to a given image, $\mathbf{x}$ is found using the lens equation

$$\mathbf{u} = \mathbf{x} - \nabla\phi(\mathbf{x}), \tag{1}$$

where $\phi$ is the two-dimensional lensing potential (see Schneider, Ehlers & Falco 1992 for a detailed discussion). We use a pixel mapping technique to move between the source and image planes. Since lensing produces multiple images of some regions of the source, it is simplest to move from the image plane to the source plane. However, the reconstruction algorithm requires us to move from the source to the image plane, since we adjust the source pixels and need to compute the resulting image to compare to the data. We therefore construct a pixel weighting function that lets us map the source plane onto the image plane without having to solve for the multiple roots of the lens equation (1).

The image plane is regarded as a grid of pixels, and the corner of each pixel is mapped back to the source plane using the lens equation. We divide each image pixel diagonally into two triangles, and each triangle is projected into a new triangle on the source plane (Blandford & Kochanek 1987, Wallington & Narayan 1993, WKK). We calculate the area of overlap, $w_{ijkl}$, of the two projected triangles for image pixel $i, j$ with each source pixel $k, l$. We use a discrete notation and adopt the convention that the indices $i$ and $j$ are used in the image plane and $k$ and $l$ in the source plane. The sum of the weights over the source plane is the inverse magnification, $\mu_{ij}$, of an image pixel,

$$\mu_{ij}^{-1} = \frac{\sum_{kl} w_{ijkl}}{\Delta x_i^2}, \tag{2}$$

where $\Delta x_i^2$ is the area of an image pixel, and the sum of weights over the image plane is the number of images $N_{kl}$ of a source pixel

$$N_{kl} = \frac{\sum_{ij} w_{ijkl}}{\Delta x_s^2}, \tag{3}$$

where $\Delta x_s^2$ is the area of a source pixel.

We compute the weights at the start of the inner cycle for each new lens model. We can then transform the array of source fluxes, $S_{kl}$, into the lensed image, $L_{ij}$, using

$$L_{ij} = \frac{\sum_{kl} w_{ijkl} S_{kl}}{\sum_{kl} w_{ijkl}}. \tag{4}$$

The image and source grids need not coincide, and we allow the program to "zoom in" on the source plane to concentrate the pixels in the most highly magnified regions of the source. The singly-imaged pixels which lie off the edge of this smaller source grid are mapped directly back to the image, as would have occurred with a larger source grid.

## 2.2. MEM Equations Including Lensing

Our algorithm to find the optimal source that fits the observed image for a fixed lens model is a generalization of the maximum entropy method, MEM, used to deconvolve the point spread function from astronomical observations (e.g. Narayan & Nityananda 1986). Given a lensed image $L$ observed by a telescope with beam $B$, the observed data $D$ is

$$D = L * B + \sigma, \tag{5}$$

where "$*$" represents a convolution and $\sigma$ is the random noise in the observation. Since many sources will fit the data to within the noise, a method such as MEM is needed to pick the "best" one. MEM accomplishes this by maximizing the entropy of the source while minimizing a measure of the error between the data and the reconstruction. This produces a map which represents the most probable source. We maximize the function

$$J = -\sum_{kl} S_{kl} \ln S_{kl} - \lambda \sum_{ij} \left( \frac{I_{ij} - D_{ij}}{\sigma_{ij}} \right)^2, \qquad (6)$$

where the first term is entropy, the second term is the $\chi^2$ difference between the data $D$ and the model image $I$, and $\lambda$ is a Lagrangian multiplier. In the case of gravitational lensing, we fit the source model, but we observe the lensed image of the source, $L$, so the model image we compare to the data is $I = L * B$. The maximum entropy equation (6) thus represents an entropy calculated on the source plane and $\chi^2$ calculated on the image plane (see WNK for a more complete discussion).

### 2.3.  Solving the MEM Equation

The goal of the maximum entropy method is to find the highest entropy solution that has a $\chi^2$ equal to the number of degrees of freedom in the map, $N_{dof}$. In practice, the target $N_{dof}$ is ambiguous because of the problems associated with using the heavily processed Clean map (see Kochanek 1995, Ellithorpe, Kochanek, & Hewitt 1995). To reach this goal we must simultaneously optimize the source and adjust the Lagrangian multiplier, $\lambda$. We follow the procedures of Cornwell & Evans (1985), and maximize $J$ using the conjugate gradient method (e.g. Press et al. 1992). The conjugate gradient method maximizes $J$ using a series of one-dimensional line maximizations in directions chosen not to interfere with previous steps. Each inner cycle with a new lens model starts with the same initial condition for $\lambda$ and a uniform source. In each iteration we produce a beam-convolved image using the current model for the source, compute the $\chi^2$ fit of this image to the data, and calculate the entropy of the current source model. We can then maximize $J$ and adjust the source model.

At the same time, we adjust the Lagrange multiplier, $\lambda$. We begin the maximization with a low value of $\lambda$, so that the entropy term dominates the value of $J$. As the maximization progresses, we increase $\lambda$ to add more information from the $\chi^2$ into the optimization. The goal is to follow the curve of the minimum value of the $\chi^2$ as a function of the Lagrange multiplier $\lambda$, starting from smooth functions that fit the data poorly and moving towards the smoothest model that fits the data well. Lengthy experience with this process has shown that the increase in $\lambda$ need not be gradual, but that $\lambda$ must only be adjusted when the algorithm has achieved some local stability and we are near the minimum value of the $\chi^2$ for the current value of $\lambda$. Thus we calculate a measure of the overall gradient of the function at each iteration and only allow $\lambda$ to be adjusted when this gradient measure is small. We also reset the conjugate gradient progression each time $\lambda$ is adjusted. This prevents the directions stored from the iterations run with the old $\lambda$ from restricting the freedom of the method to find the best solution with the new $\lambda$.

The other important consideration in carrying out the maximization is choosing a stopping criterion. Simply stopping the inversion when the $\chi^2$ reaches some target value such as $N_{pix}$, the number of pixels in the map, is impractical in the limit of both very good and very bad models. With a good model, the final $\chi^2$ can typically be far smaller than $N_{pix}$, whereas with a bad model the MEM reconstruction will not be able to fit the data no matter how long it runs. In both cases, using a target value of $\chi^2$ as a stopping criterion would be counter-productive. For that reason we have put a simple iteration requirement into our program: every MEM reconstruction runs for 100 iterations. We find that this allows for super-resolution in those models which can achieve it, and that those models which cannot reach an acceptable $\chi^2$ level after 100 iterations probably never will.

We define error bars for our lens parameters by performing a series of Monte Carlo simulations. In this way we avoid some of the vagaries encountered by Kochanek & Narayan (1992), WKK, Kochanek (1995), and Ellithorpe, Kochanek & Hewitt (1995) in trying to define error bars from the value of $\chi^2$. Because of the finite size of the beam, adjoining pixels are not independent, and so the number of degrees of freedom in the map is less than $N_{pix}$. In addition, poor lens models will produce residuals in the multiply imaged portion of the map which are significantly higher than those found in the single image region. Monte Carlo techniques allow us to avoid these difficulties and still define reliable error bars for our lens solutions.

## 2.4. Lens Potential Model

We use a simple lens potential model with only five parameters. We model the lens as a singular isothermal sphere embedded in an external shear field,

$$\Phi = br + \frac{1}{2}\gamma r^2 \cos 2(\theta - \theta_\gamma),\tag{7}$$

where $b = 4\pi(\sigma/c)^2 D_{ds}/D_{os}$ is the critical radius of the lens, $\gamma$ is the dimensionless shear, $\theta_\gamma$ is the angle of the shear, $\sigma$ is the one dimensional velocity dispersion, and $D_{ds}$ and $D_{os}$ are the angular diameter distances between the lens and the source and observer and source, respectively. Potentials of this form can roughly fit all the observed cases of lensed quasars (e.g. Kochanek 1991). The five parameters of this model are the coordinates of the lens center, $x_l, y_l$, the critical radius, $b$, the shear, $\gamma$, and the shear angle, $\theta_\gamma$. The outer cycle of the algorithm optimizes these lens parameters using simplex minimization (e.g. Press et al. 1992) of the final $\chi^2$ produced by the inner cycle MEM reconstruction.

## 3. Tests of the Algorithm

We performed a series of tests of the algorithm using artificial data. By inverting a known image we can test how well our algorithm solves for the lens parameters and generates a good source model.

We began with the source model shown in Figure 1a. We purposely used a source model that resembles the real lens reconstruction of MG1654+134 (Kochanek 1995), so we could test how well our algorithm models low-lying details and complicated structure on scales smaller than the beam. We used the lens parameters listed under "Real Values" in Table 1 to produce the image, and convolved the model with a Gaussian beam with FWHM equal to 5 pixels. We added Gaussian random noise to the image so that the peak signal-to-noise ratio in the image is S/N=150. Figure 1b shows the "data" that we constructed from this source and lens model. The source plane grid is 128 by 128 pixels and the image is 64 by 64 pixels. The source has also been "zoomed" in by a factor 1.88 with respect to the image plane, making each source pixel 0.266 as big as the image pixels.

We produced the lensed "data" analytically rather than with our pixel mapping technique. Since the same pixel weights would be used to invert the lens, using them to create the initial image would eliminate the numerical inaccuracies of the mapping algorithm and misrepresent the accuracy of the method. To produce the initial data image we divided each image pixel into 25 subpixels, found the source flux corresponding to each subpixel center using a 4-point interpolation scheme, and then averaged over all the subpixels.

### 3.1. Lens Parameter Solution

Our first test of the method was to compute the $\chi^2$ as a function of a single lens parameter while holding the other four fixed at their true values. We repeated this test for several different noise realizations. This gives us a sense of the error bars on each parameter and lets us check the smoothness of the $\chi^2$ in each dimension of the parameter space before doing the full five-parameter minimization. In the case of real lens data we might start the inversion by constraining the lens parameters in this way to develop a feel for the parameter values and sensitivities. Figure 2 shows the $\chi^2$ of the residuals as a function of $x_l$, $b$, and $\gamma \cos \theta_\gamma$, with curves for five different noise realizations. The minimum for each curve is near the real value of that parameter, and the $\chi^2$ level is well below 4096, the number of pixels in the data. The $\chi^2$ is able to reach this low value because some of the degrees of freedom of the model are used to build the source.

The second test was to perform minimizations in which three parameters, $b$, $\gamma$, and $\theta_\gamma$, were allowed to vary while the remaining two parameters, $x_l$ and $y_l$ were held fixed. Figure 3 shows a contour plot of the final $\chi^2$ values for these minimizations. Again, we find a well constrained minimum, and the values of $b$, $\gamma$ and $\theta_\gamma$ are close to what we know to be the real values.

The final test was a full five parameter minimization. Table 1 shows the results of 25 five parameter minimizations using different noise realizations. We used the standard deviation of the calculated parameters to derive error bars for the solution. The average value of each parameter lies within two standard deviations of the real value for every parameter except $\gamma$. Figure 4 shows scatter plots of parameters pairs. Two of the pairs of parameters, $\gamma$ vs. $\theta_\gamma$ and $y_l$ vs. $\gamma$, seem to be correlated, but the errors in these parameters are still small.

### 3.2. Source Reconstruction

In addition to providing a reliable lens model, a good inversion algorithm should be able to reproduce the source. Figure 5b shows the source reconstruction of an image made with a beam equal to our original beam. We analyzed the source reconstructions of a series of image models in which we varied the signal-to-noise ratio and the resolution of the image. Unlike similar tests in WNK, the signal-to-noise ratio had little effect on the accuracy of the source reconstruction. The resolution, however, had a significant impact. For images created using beams much larger than that used in the original model (FWHM = 10 and 20 pixels, vs. the original FWHM = 5 pixels), the algorithm was unable to produce a reasonable inversion. These simulations had final $\chi^2$ values more than ten times higher than those found in our other inversions. For images created using a beam half the size of our original value, the image reconstruction fit the data very well, but the source reconstruction was highly fragmented (Figure 5a).

As discussed in WNK, many deconvolution routines produce reconstructions with too much high frequency power, but the extra power can be filtered out by smoothing with a suitably selected beam. In the Clean algorithm, for instance, the source is reconstructed as an ensemble of delta functions, which are smoothed to produce the source map. Figures 5c and 5d show the results of smoothing the source reconstructions with a gaussian beam with FWHM= 2.5. These smoothed sources more accurately recreate the true source.

### 3.3. Source Reconstruction Glitches

In WNK, where we first formulated the LensMEM equations in one dimension, we spent a considerable amount of effort investigating "glitches" which appeared in the reconstruction of large sources. These spiky patterns appeared at the boundaries between regions of different image multiplicity in sources which were large compared to the lens critical radius. We developed a second formulation of the LensMEM equations to try to combat the problem, but found the glitches to be intrinsic to the maximum entropy method when combined with lensing. It was not clear how much of an effect these glitches would have on the full two-dimensional algorithm. Now that we have the complete algorithm, we reproduced the conditions that led to the glitches with our artificial data. We were not able to produce any obvious glitches in two dimensions with either formulation of the MEM equations. It is possible that the averaging produced by overlapping pixels has diluted the glitch effect, which was quite pronounced in one dimension.

### 4. Inversion of MG1654+134

Since LensMEM works on artificial data, we next attempted to model a real lens, the radio ring MG1654+134. Recent detailed models of this system by Kochanek (1995) and Ellithorpe, Kochanek & Hewitt (1995) using the LensClean algorithm provide an interesting opportunity to compare the results obtained using these complementary methods.

Figure 6a shows the 8 GHz VLA image of MG1654+134 from Langston et al. (1990). This $128 \times 128$ image has a circular Clean beam with a FWHM of $0\rlap{.}''18$ (six pixels), a signal to noise ratio of 134, and a pixel scale of $0\rlap{.}''03$. The source is at redshift $z_s = 1.74$ and the lens is at $z_l = 0.254$.

We model the lens using a singular isothermal sphere embedded in an external shear field (equation 8) and adopt a source plane grid with $128^2$ pixels on the same scale as the image plane (i.e. we did not "zoom in"). We first performed a series of three-parameter optimizations in a grid of fixed lens positions $x_l, y_l$. Table 2 shows the $\chi^2$ values found at each position. Since the position and parameter values were in rough agreement with the position found by Kochanek (1995), we decided to do a full five-parameter minimization after just this preliminary test.

The first line of Table 3 shows the lens parameters found with a five parameter minimization. The final $\chi^2$ of this model was 9081, which is 0.55 per image pixel. In order to obtain error bars for the individual lens parameters, we used our best lens model to produce a reconstructed image (Figure 6b) of MG1654+134 and added random noise with the same S/N to generate model data. We then performed a series of five parameter reoptimizations of the model using the reconstructed data with different noise realizations. Table 3 shows the results of these optimizations. The average $\chi^2$ found for the reconstructed data was considerably below that found for the real data, since the noise in the reconstructed data is truly random. The real data includes systematic errors, since it is a Clean image constructed from visibility data. The ideal way to carry out the Monte Carlo simulations would be to start from the visibilities.

We also compare our results with those found by Kochanek (1995), which are also listed in Table 3. All our parameter values are well within one standard deviation of the Kochanek (1995) values except for the lens position, in which we disagree by approximately $1.3\sigma$.

Figure 6c shows our best source reconstruction (obtained by inverting the real data with our best lens model), smoothed with an appropriate beam. This source is qualitatively the same as that obtained by Kochanek (1995). Figure 6d shows the residuals of the reconstructed image.

## 5. Conclusions

We presented a new lens inversion algorithm, LensMEM, that uses the maximum entropy method to account for the finite resolution in lens observations. We are able to simultaneously produce a model source and lens using the constraints from multiple imaging. We tested the algorithm using artificial data and Monte Carlo simulations, and showed that it produces reliable lens parameters. The parameter values are consistent with the true values, and the $\chi^2$ errors around the minima have a smooth, quadratic form. These $\chi^2$ values are low compared with the number of pixels in the image because many degrees of freedom are used to build the model of the source. We estimate parameter errors using Monte Carlo simulations of the data with different noise realizations in which we optimize the lens model and calculate the standard deviation of the solutions. We also find that the algorithm produces good models for the source if the image is sufficiently well resolved. Very low resolution maps cannot be properly inverted, but high resolution maps fit the data well while producing fragmented source reconstructions. Smoothing by a suitable beam eliminates the high frequency structure and gives a good source reconstruction.

We applied the algorithm to real data by modeling the radio ring MG1654+134. We compared our lens parameter solutions with those found by Kochanek (1995) using the comparable inversion algorithm LensClean. The largest deviation from Kochanek (1995) was $1.3\sigma$. We again performed Monte Carlo simulations to calculate error bars for our parameters by adding random noise to our reconstruction of the data. The $\chi^2$ fits of the Monte Carlo inversions were much lower than the reconstruction performed on the real data, since we were able to insure truly random noise in our reconstructed data. A possible additional improvement to the algorithm would be to modify it to operate directly on radio visibility data (Ellithorpe, Kochanek & Hewitt 1995). This would circumvent the systematic errors introduced by Clean in producing the image.

Lensing has become a very active field, with new lensed objects constantly being discovered. As the sample of lenses grows, the need for rigorous lens inversion algorithms increases. Algorithms such as LensClean and LensMEM are vital for fully exploiting gravitational lenses. We have shown LensMEM to be successful at both producing good source maps and obtaining reliable lens parameters, and we hope to see it become a valuable tool in the study of gravitational lensing.

Acknowledgements: This work was supported in part by the NSF through grants AST-9109525 (RN) and AST-9401722 (CSK).

| Table 1: Five-Parameter Solutions for Artificial Data Inversion | | | | | | |
|---|---|---|---|---|---|---|
| Noise Realization | $x_l$ | $y_l$ | $b$ | $\gamma$ | $\theta_\gamma$ | $\chi^2$ |
| 1 | 31.898 | 30.138 | 10.079 | 0.0460 | 104.°2 | 1569.0 |
| 2 | 31.874 | 29.999 | 10.092 | 0.0487 | 102.0 | 1776.8 |
| 3 | 31.673 | 30.071 | 10.141 | 0.0493 | 102.3 | 1420.2 |
| 4 | 31.862 | 30.011 | 10.087 | 0.0484 | 102.2 | 1794.8 |
| 5 | 31.972 | 30.203 | 10.081 | 0.0450 | 105.2 | 1613.8 |
| 6 | 31.878 | 30.051 | 10.088 | 0.0478 | 102.8 | 1443.4 |
| 7 | 31.797 | 30.116 | 10.093 | 0.0470 | 103.5 | 1597.3 |
| 8 | 31.803 | 30.094 | 10.092 | 0.0473 | 103.4 | 1550.3 |
| 9 | 31.906 | 30.047 | 10.094 | 0.0477 | 102.7 | 1393.5 |
| 10 | 31.662 | 30.072 | 10.102 | 0.0477 | 103.5 | 1475.9 |
| 11 | 31.607 | 30.061 | 10.089 | 0.0474 | 103.5 | 1656.6 |
| 12 | 31.607 | 30.061 | 10.089 | 0.0474 | 103.5 | 1520.0 |
| 13 | 31.855 | 30.160 | 10.091 | 0.0461 | 104.3 | 1568.3 |
| 14 | 31.649 | 30.114 | 10.090 | 0.0477 | 103.7 | 1553.1 |
| 15 | 31.865 | 30.139 | 10.091 | 0.0475 | 103.7 | 1803.8 |
| 16 | 31.895 | 30.134 | 10.079 | 0.0466 | 104.1 | 1618.8 |
| 17 | 31.842 | 30.113 | 10.092 | 0.0471 | 103.4 | 1572.7 |
| 18 | 31.837 | 30.105 | 10.073 | 0.0465 | 104.0 | 1682.3 |
| 19 | 31.688 | 30.074 | 10.091 | 0.0479 | 103.3 | 1731.0 |
| 20 | 31.767 | 30.024 | 10.106 | 0.0493 | 102.1 | 1754.4 |
| 21 | 31.978 | 30.122 | 10.075 | 0.0468 | 103.1 | 1705.4 |
| 22 | 31.732 | 30.100 | 10.090 | 0.0469 | 103.8 | 1664.6 |
| 23 | 31.688 | 30.031 | 10.142 | 0.0475 | 103.4 | 1645.0 |
| 24 | 31.688 | 30.135 | 10.089 | 0.0476 | 103.4 | 1713.4 |
| 25 | 31.708 | 30.112 | 10.082 | 0.0471 | 103.9 | 1621.2 |
| Average | 31.789 | 30.091 | 10.093 | 0.0474 | 103.4 | 1617.8 |
| rms | 0.109 | 0.048 | 0.016 | 0.0010 | 0.7 | 112.0 |
| Real Values | 32.000 | 30.100 | 10.079 | 0.0453 | 102.9 | |
| Fit of Monte Carlo Solution | $1.94\sigma$ | $0.19\sigma$ | $0.88\sigma$ | $2.10\sigma$ | $0.71\sigma$ | |

Notes: This table gives the results of 25 different LensMEM reconstructions of the artificial data shown in Figure 1. The columns are the lens position, $x_l, y_l$ and the bending angle, $b$, in pixels, the shear, $\gamma$, and the orientation of the shear, $\theta_\gamma$, in degrees. The final column shows the $\chi^2$ fit of the reconstructed image to the data. The average values of the lens parameters are given at the bottom of the table, with their rms error bars. The last line shows the deviation of the average values from the true values in units of the estimated uncertainties.

| Table 2: $\chi^2$ Values for a Grid of Lens Positions | | | | |
|---|---|---|---|---|
| $y_l$ ⟍ $x_l$ | 1″775 | 1″805 | 1″835 | 1″865 |
| 1″848 | 92185.4 | 87090.4 | 62823.2 | 108887.2 |
| 1″878 | 25518.3 | 17320.6 | 19008.4 | 29621.9 |
| 1″908 | 32727.1 | 23424.2 | 11393.4 | 22750.2 |
| 1″938 | 99711.5 | 87973.9 | 80629.2 | 77801.2 |

Notes: This table gives the final $\chi^2$ for three-parameter optimizations of MG1654+134 made with fixed lens positions.

| Table 3: Five-Parameter Solutions for MG1654+134 Inversion | | | | | | |
|---|---|---|---|---|---|---|
| | $x_l$ | $y_l$ | $b$ | $\gamma$ | $\theta_\gamma$ | $\chi^2$ |
| Original Data Solution | 1″8251 | 1″8896 | 0″9820 | 0.0771 | 100°8 | 9081.9 |
| Noise Realization 1 | 1.8264 | 1.8911 | 0.9825 | 0.0775 | 101.1 | 5796.6 |
| 2 | 1.8254 | 1.8906 | 0.9825 | 0.0768 | 100.9 | 6447.1 |
| 3 | 1.8235 | 1.8940 | 0.9817 | 0.0784 | 101.2 | 4433.7 |
| 4 | 1.8261 | 1.8884 | 0.9835 | 0.0757 | 100.9 | 4885.9 |
| 5 | 1.8260 | 1.8903 | 0.9817 | 0.0776 | 101.0 | 5479.7 |
| 6 | 1.8239 | 1.8908 | 0.9808 | 0.0779 | 101.2 | 5919.4 |
| 7 | 1.8246 | 1.8885 | 0.9806 | 0.0769 | 101.3 | 4856.2 |
| 8 | 1.8270 | 1.8888 | 0.9821 | 0.0772 | 100.8 | 5083.7 |
| 9 | 1.8260 | 1.8906 | 0.9820 | 0.0773 | 101.0 | 4937.7 |
| 10 | 1.8273 | 1.8926 | 0.9813 | 0.0793 | 101.5 | 5128.3 |
| 11 | 1.8285 | 1.8891 | 0.9827 | 0.0773 | 100.7 | 6451.6 |
| 12 | 1.8241 | 1.8898 | 0.9824 | 0.0769 | 100.8 | 5600.3 |
| 13 | 1.8259 | 1.8902 | 0.9824 | 0.0773 | 101.4 | 5152.7 |
| 14 | 1.8257 | 1.8870 | 0.9816 | 0.0764 | 100.7 | 5902.5 |
| 15 | 1.8271 | 1.8911 | 0.9832 | 0.0776 | 101.0 | 5038.9 |
| Average | 1.8258 | 1.8902 | 0.9821 | 0.0773 | 101.0 | 5407.6 |
| rms | 0.0013 | 0.0017 | 0.0008 | 0.0008 | 0.2 | 577.4 |
| Kochanek (1995) Values | 1.8049 | 1.8780 | 0.9820 | 0.0770 | 100.6 | |
| rms | 0.016 | 0.012 | 0.003 | 0.005 | 1.0 | |

Notes: This table shows LensMEM reconstructions of the radio ring MG1654+134. The first line gives the results of the solution obtained from the real data. The following 15 lines represent the solutions found by inverting the reconstructed data, to which random noise was added. This allowed us to find rms errors for the solved parameters. These error bars are given with the average of the different noise realizations. The last two lines show the lens parameters and errors found by Kochanek (1995). The lens positions reflect a coordinate transformation between the our respective models. The column headings are the same as in Table 1.

---

## Figure Captions

Fig. 1—Artificial data. We project an artificial source (a) through a lens with parameters given in Table 1, smooth it with a beam, and add random noise to produce the "data" (b). The source contains $128^2$ pixels and the image has $64^2$ pixels. The different scales for the source and image planes reflect the fact that the source is "zoomed-in" with respect to the image. The beam has a FWHM = 5 pixels and the peak signal to noise in the image is S/N=150.

Fig. 2—One-dimensional cuts through the error surface, made by varying one parameter while holding the other four fixed. In (a) the lens position $x_l$ is varied, in (b) the bending angle $b$ is varied, and in (c) the quantity $\gamma \cos \theta_\gamma$ is varied. In the latter case the parameters held fixed are $x_l, y_l, b$ and $\gamma \sin \theta_\gamma$. Six different Monte Carlo simulations are shown in each case.

Fig. 3—Contours of final $\chi^2$ for three-parameter optimizations in an array of lens positions, $x_l, y_l$. The $\chi^2$ shows a minimum which is close to the true value, indicated by a '+'. The contours have a spacing of $\Delta \chi^2 = 300$, with the lowest contour at $\chi^2 = 1800$.

Fig. 4—Scatter plots of the optimized parameters found from inverting artificial data with 25 different noise realizations. The lens parameters, $x_l, y_l, b, \gamma$, and $\theta$ are paired together to give a sense of the correlation between solved parameters. Only panels (a) and (c) show any evidence of correlation, and the spread in these parameters is very small.

Fig. 5—Montage of source reconstructions. The same lens and source are used in each case, but the beam FWHM is varied. (a) and (b) show raw source reconstructions made with beam FWHM = 2.5 and 5 respectively. These reconstructions are then smoothed with a FWHM = 2.5 beam [(c) and (d)]. The contour levels are the same as in Figure 1, but the scale of the plots has been increased to give a better picture of the details of the sources.

Fig. 6—Reconstruction of radio ring MG1654+134. (a) 8 GHz VLA observation. (b) The LensMEM reconstructed image. (c) The smoothed source reconstruction. The butterfly-shaped boundary shows the projection of the edges of image-plane.(d) Residuals of the inversion.

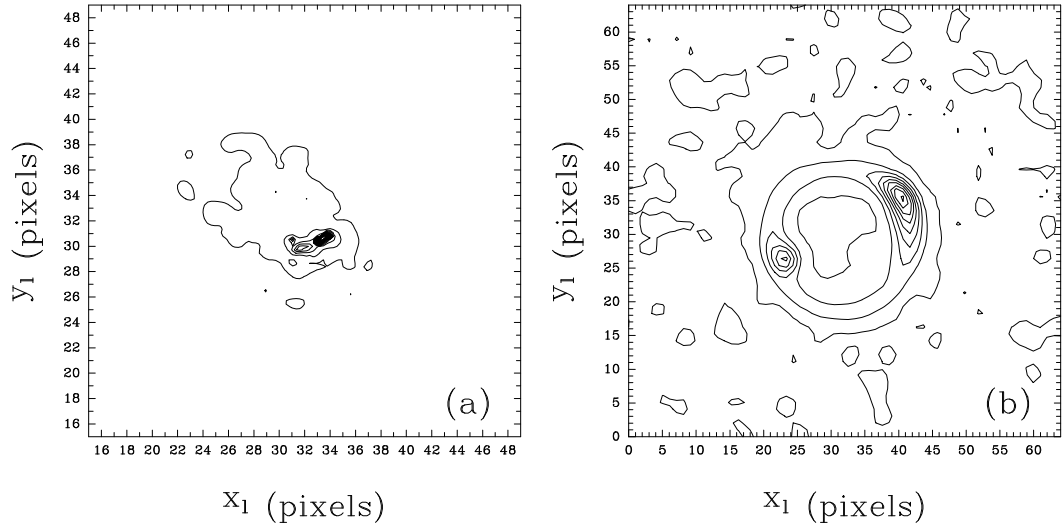

Fig. 1.—

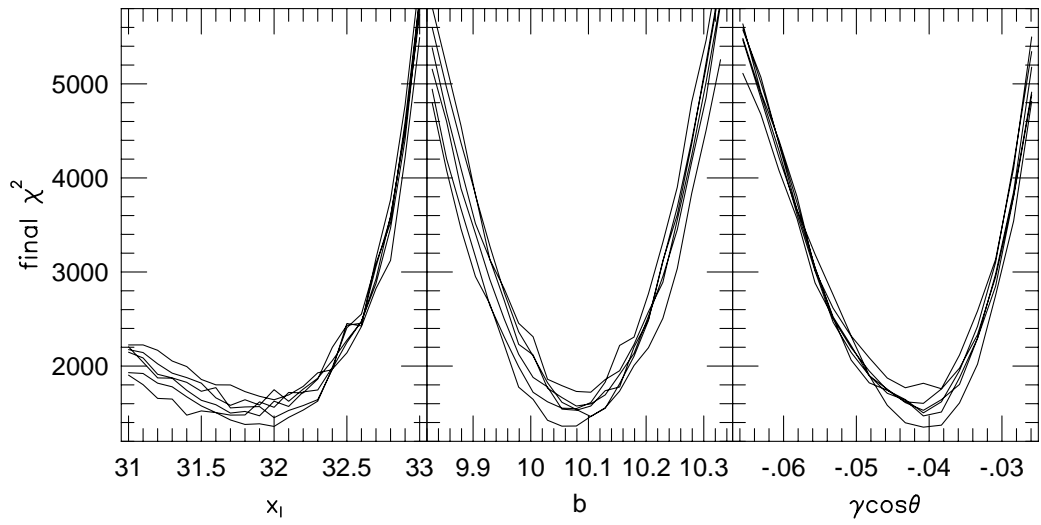

Fig. 2.—

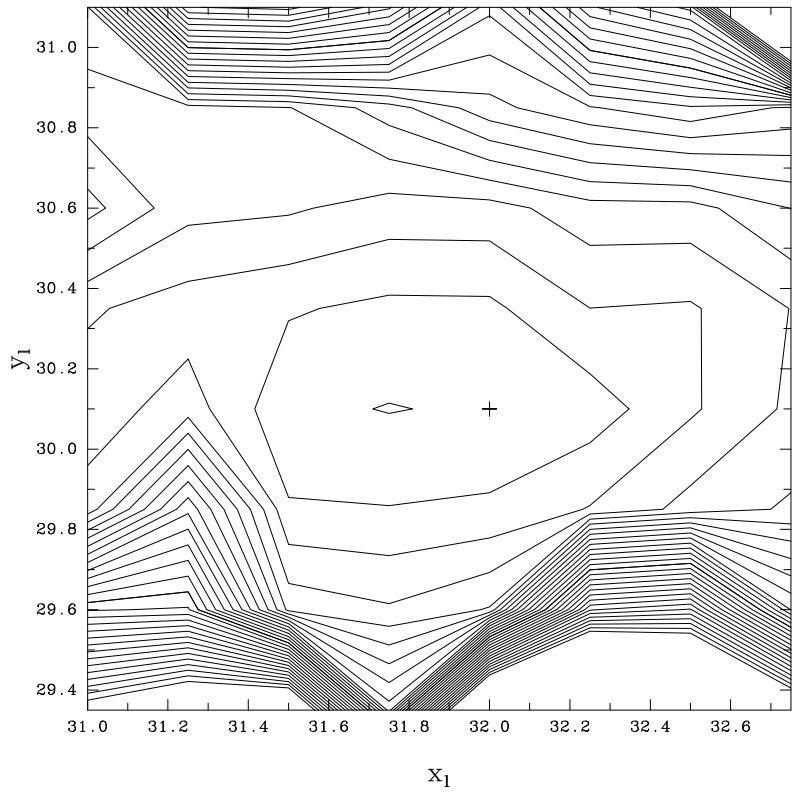

Fig. 3.—

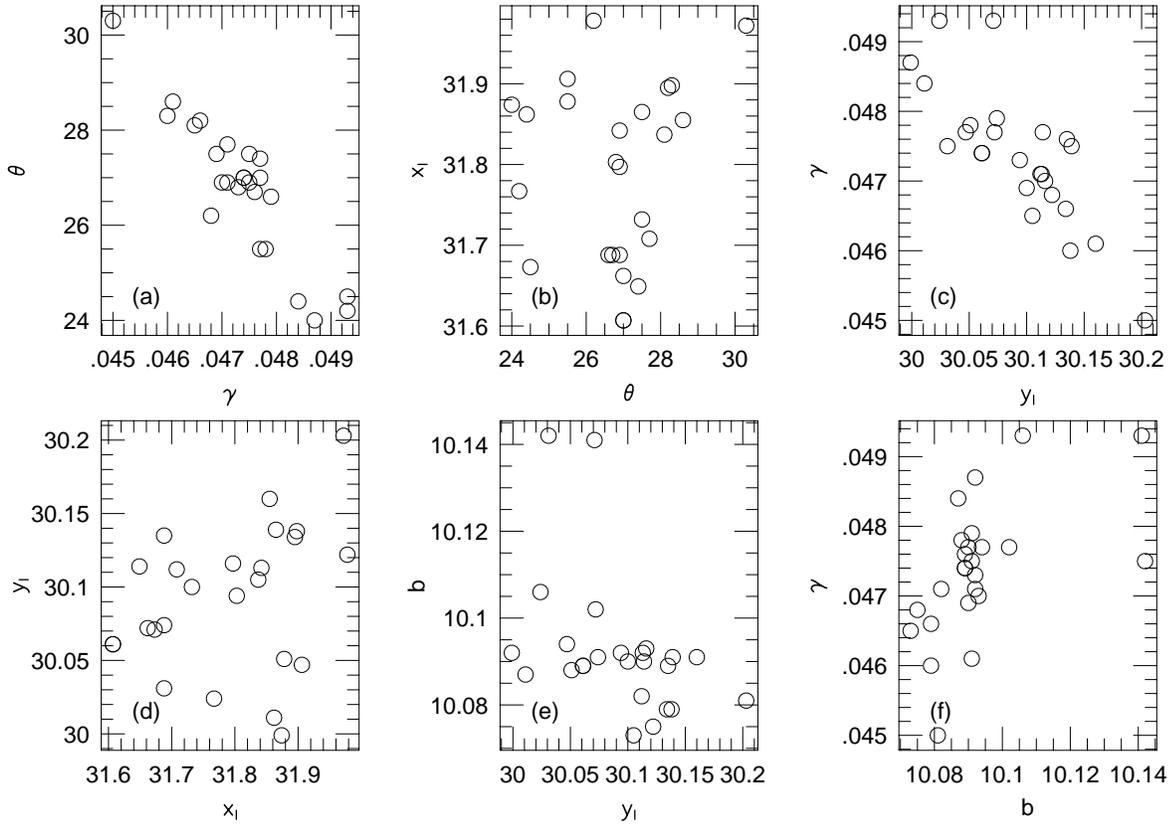

Fig. 4.—

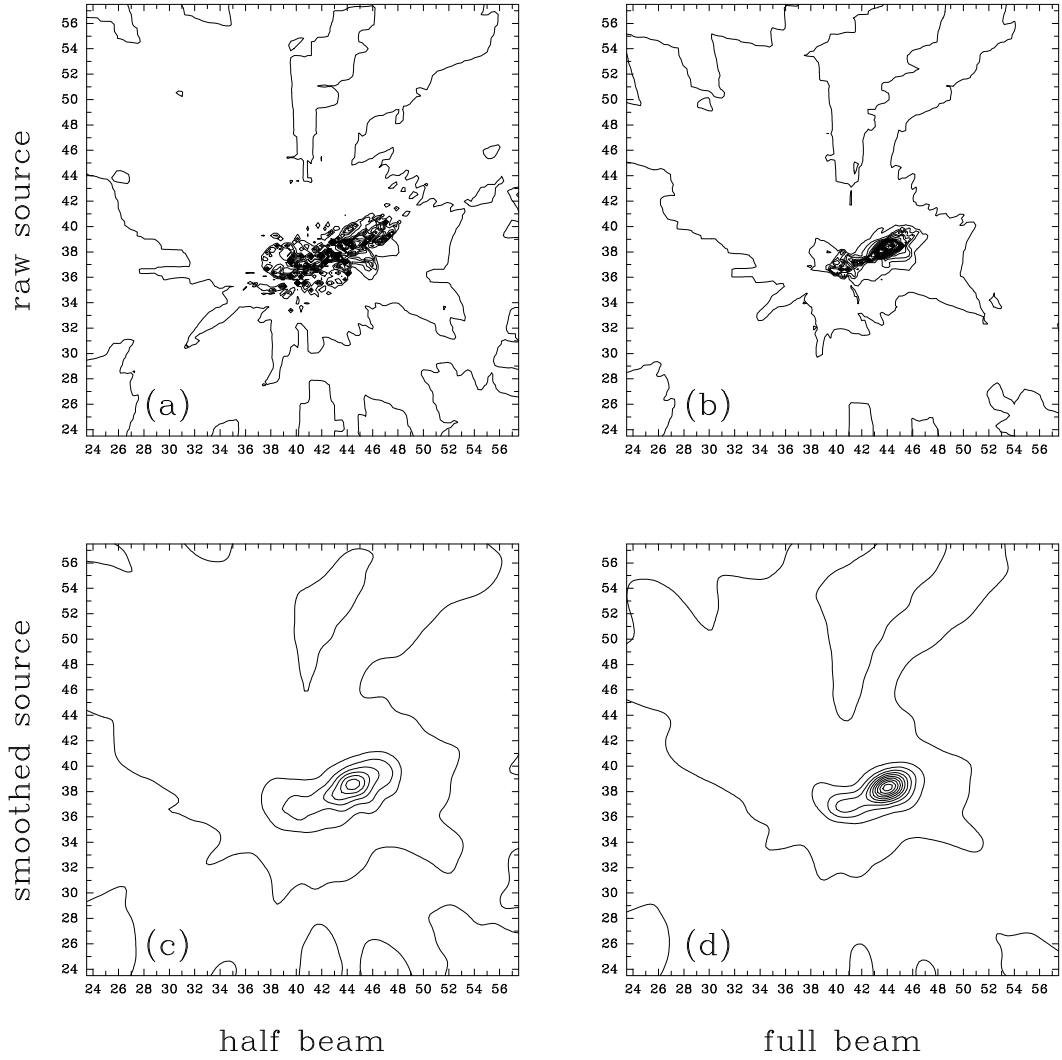

Fig. 5.—

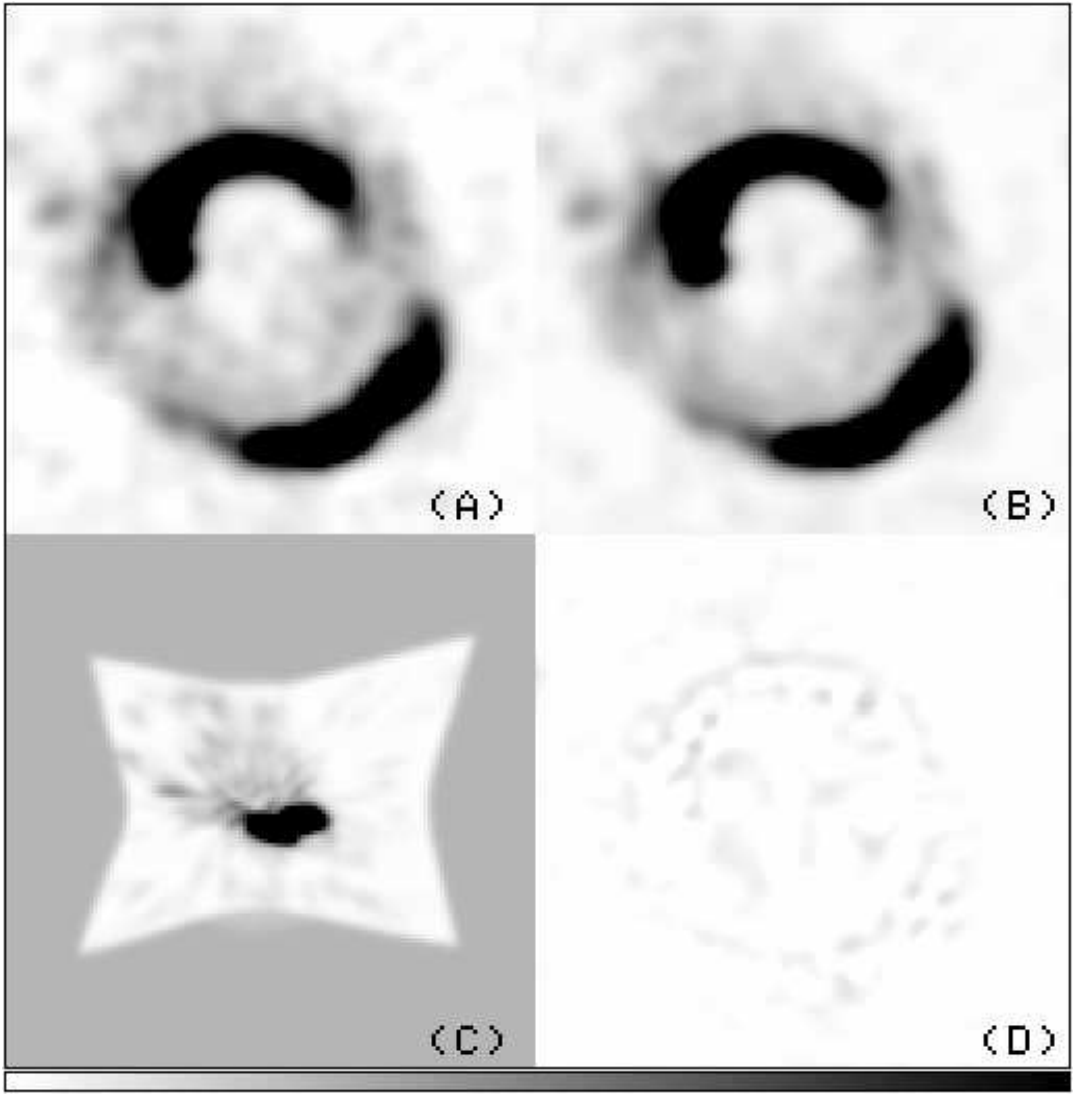

Fig. 6.—